\documentclass{appolb}
\usepackage{epsfig}
\begin{document}
\title{Strange and charm mesons at FAIR%
\thanks{Presented at XXXI Mazurian Lakes Conference on Physics: Nuclear Physics and the Road to FAIR, August 30- September 6, 2009}%
}
\author{L. Tolos$^{1}$, D. Cabrera$^{2}$, D. Gamermann$^{3}$, C. Garcia-Recio$^4$, R. Molina$^3$, J. Nieves$^3$, E. Oset$^3$ and A. Ramos$^5$
\address{$^1$Theory Group. KVI. University of Groningen, \\
Zernikelaan 25, 9747 AA Groningen, The Netherlands}\\
\address{$^2$Departamento de F\'{\i}sica Te\'orica II, Universidad Complutense,\\
28040 Madrid, Spain}\\
\address{$^3$Instituto de F{\'\i}sica Corpuscular (centro mixto CSIC-UV)\\
Institutos de Investigaci\'on de Paterna, Aptdo. 22085, 46071, Valencia, Spain}\\
\address{$^4$ Departamento de F{\'\i}sica At\'omica, Molecular y Nuclear, \\
Universidad de Granada, E-18071 Granada, Spain}\\
\address{$^5$Departament d'Estructura i Constituents de la Mat\`eria,\\
Universitat de Barcelona,
Diagonal 647, 08028 Barcelona, Spain}
}
\maketitle
\begin{abstract}
We study the properties of strange and charm mesons in hot and dense matter within a self-consistent coupled-channel approach for the experimental conditions of density and temperature expected for the CBM experiment at FAIR/GSI. The in-medium solution at finite temperature accounts for Pauli blocking effects, mean-field binding of all the baryons involved, and meson self-energies. We analyze the behaviour in this hot and dense environment of dynamically-generated baryonic resonances together with the evolution with density and temperature of the strange and open-charm meson spectral functions. We test the spectral functions for strange mesons using energy-weighted sum rules and finally discuss the implications of the properties of charm mesons on the  $D_{s0}(2317)$ and the predicted $X(3700)$ scalar resonances.
\end{abstract}
  
\section{Introduction}

Over the last decades strangeness has been a matter of extensive study in connection to exotic atoms \cite{Friedman:2007zz} as well as heavy-ion collisions at SIS/GSI energies \cite{Fuchs:2005zg}. Phenomenology of antikaonic atoms shows that the $\bar K$ feels an attractive potential at low densities. 
This attraction results from the modified $s$-wave $\Lambda(1405)$ resonance in the medium due to Pauli blocking effects \cite{Koch} together with the self-consistent consideration of the $\bar K$ self-energy \cite{Lutz} and the inclusion of self-energies of the mesons and baryons in the intermediate states \cite{Ramos:1999ku}. Attraction of the order of -50 MeV at normal nuclear matter density, $\rho_0=0.17 \,{\rm fm^{-3}}$, is obtained by different approaches, such as unitarizated theories in coupled channels based on meson-exchange models \cite{Tolos01} or chiral dynamics \cite{Ramos:1999ku}. Higher-partial waves beyond the s-wave contribution have been also studied \cite{Tolos:2006ny,Lutz:2007bh,Tolos:2008di} as they become relevant for heavy-ion collisions at beam energies below 2AGeV \cite{Fuchs:2005zg}.

Also the charm degree of freedom is a recent topic of analysis in heavy-ion experiments. The CBM experiment of the future FAIR project at GSI will investigate highly compressed dense matter in nuclear collisions with a beam energy range between 10 and 40 GeV/u. An important part of the hadron physics project is devoted to extend the SIS/GSI program for the in-medium modification of hadrons to the heavy quark sector providing a first insight into charm-nucleus interaction. Thus, the possible modifications of the properties of open and hidden charm mesons in a hot and dense environment are matter of recent studies.

The in-medium modification of the open charm mesons ($D$ and $\bar D)$ may help to explain the $J/\Psi$ suppression in a hadronic environment as well as the possible formation of $D$-mesic nuclei. Moreover, changes in the properties of open charm mesons will affect the renormalization of charm and hidden charm scalar meson resonances in nuclear matter, providing information about their nature, whether they are $q\bar{q}$ states, molecules, mixtures of $q\bar{q}$ with meson-meson components, or dynamically generated resonances resulting from the interaction of two pseudoscalars.

In the present article, we present a study of the properties of strange and charm mesons in hot and dense matter within a self-consistent approach in coupled channels for the conditions expected at CBM/FAIR. We analyze the behaviour of dynamically generated baryonic resonances as well as the strange and charm meson spectral functions in this hot and dense medium. We then test our results for strange mesons using energy-weighted sum rules and analyze the effect of the self-energy of $D$ mesons on dynamically-generated charm and hidden charm scalar resonances.

\section{Strange and charm mesons in hot and dense matter}

The self-energy and, hence, the spectral function at finite temperature for strange ($\bar K$ and $K$) and charm ($D$ and $\bar D$) mesons are obtained following a self-consistent coupled-channel procedure. We start by solving the Bethe-Salpeter equation in coupled channels or $T$-matrix ($T$)  taking, as bare interaction, a transition potential coming from effective lagrangians. Details about this bare interaction for strange and charm mesons are given in the next sections. The self-energy is then obtained summing the transition amplitude $T$ for the different isospins over the nucleon Fermi distribution at a given temperature, $n(\vec{q},T)$, as 
\begin{eqnarray}
\Pi(q_0,{\vec q},T)= \int \frac{d^3p}{(2\pi)^3}\, n(\vec{p},T) \,
[\, {T}^{(I=0)} (P_0,\vec{P},T) +
3 \, {T}^{(I=1)} (P_0,\vec{P},T)\, ]\ , \label{eq:selfd}
\end{eqnarray}
where $P_0=q_0+E_N(\vec{p},T)$ and $\vec{P}=\vec{q}+\vec{p}$ are
the total energy and momentum of the meson-nucleon pair in the nuclear
matter rest frame, and ($q_0$,$\vec{q}\,$) and ($E_N$,$\vec{p}$\,) stand  for
the energy and momentum of the meson and nucleon, respectively, also in this
frame. The self-energy must be determined self-consistently since it is obtained from the
in-medium amplitude $T$ which contains the meson-baryon loop function, and this last quantity itself
is a function of the self-energy. The meson spectral function then reads
\begin{equation}
S(q_0,{\vec q}, T)= -\frac{1}{\pi}\frac{{\rm Im}\, \Pi(q_0,\vec{q},T)}{\mid
q_0^2-\vec{q}\,^2-m^2- \Pi(q_0,\vec{q},T) \mid^2 } \ .
\label{eq:spec}
\end{equation}

\section{Strange mesons}

The kaon self-energies in symmetric nuclear matter at finite
temperature are obtained from the in-medium kaon-nucleon interaction within a chiral unitary approach. The model incorporates the 
$s$- and $p$-waves of the kaon-nucleon interaction \cite{Tolos:2008di}. 

At tree level, the $s$-wave amplitude arises from the Weinberg-Tomozawa (WT) term of the chiral Lagrangian.
Unitarization in coupled channels is imposed by solving the Bethe-Salpeter
equation with on-shell amplitudes and a cutoff regularization. The
unitarized $\bar K N$ amplitude generates dynamically the $\Lambda(1405)$
resonance in the $I=0$ channel and provides a satisfactory description of
low-energy scattering observables. The in-medium solution of the $s$-wave amplitude accounts for Pauli-blocking
effects, mean-field binding on the nucleons and hyperons via a $\sigma-\omega$
model, and the dressing of the pion and kaon propagators via their corresponding
self-energies in a self-consistent manner. The model includes, in addition, a $p$-wave contribution to
the self-energy from hyperon-hole ($Yh$) excitations, where $Y$ stands for $\Lambda$, $\Sigma$ and
$\Sigma^*$ components.

%%%%%%%%%%%%%%%%%%%%%%%%%%%%%%%%%%%%%%%%%%%%%%%%%%%
\begin{figure}[t]
\begin{center}
\includegraphics[height=7 cm, width=6 cm]{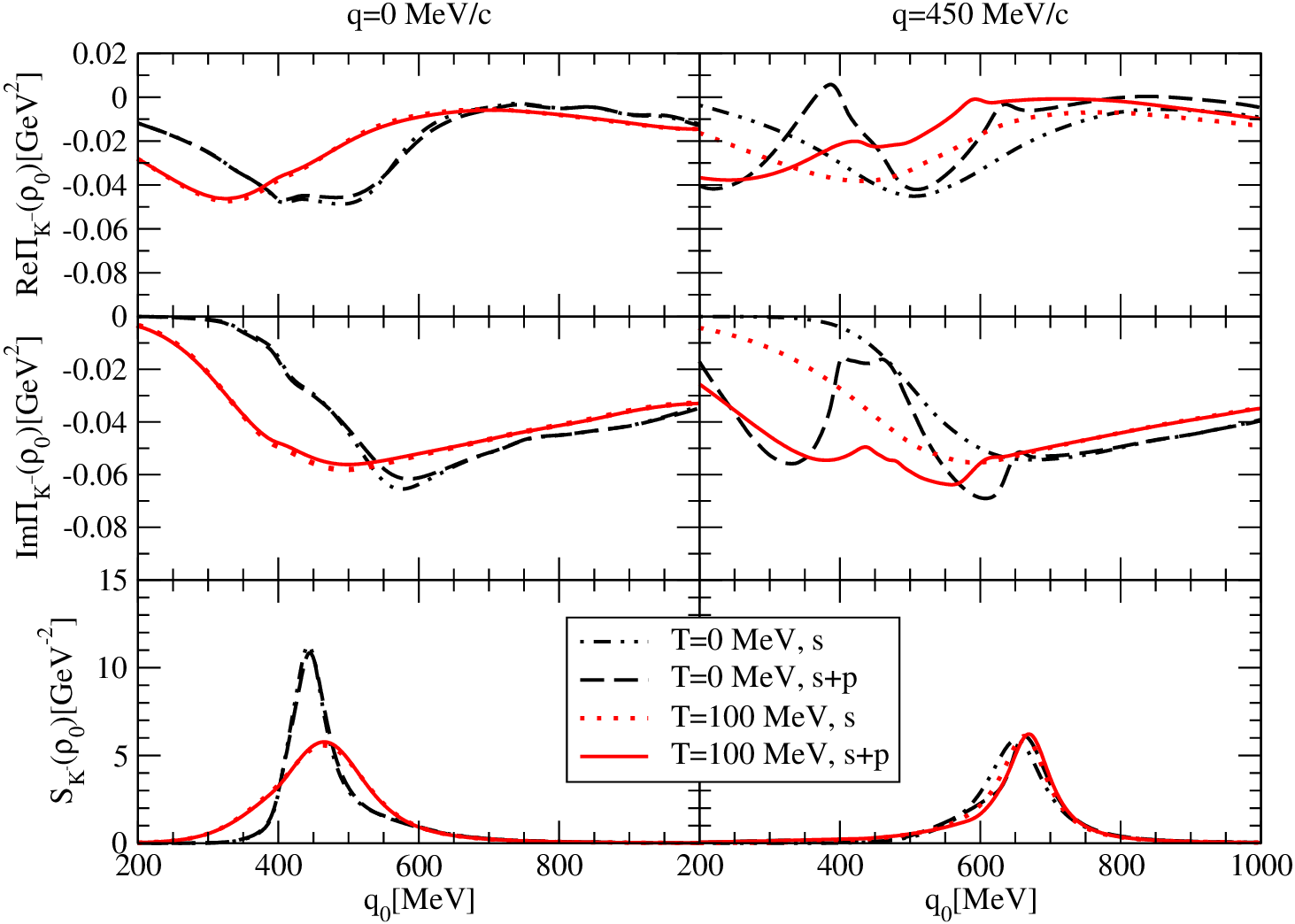}
\hfill
\includegraphics[height=7 cm, width=6 cm]{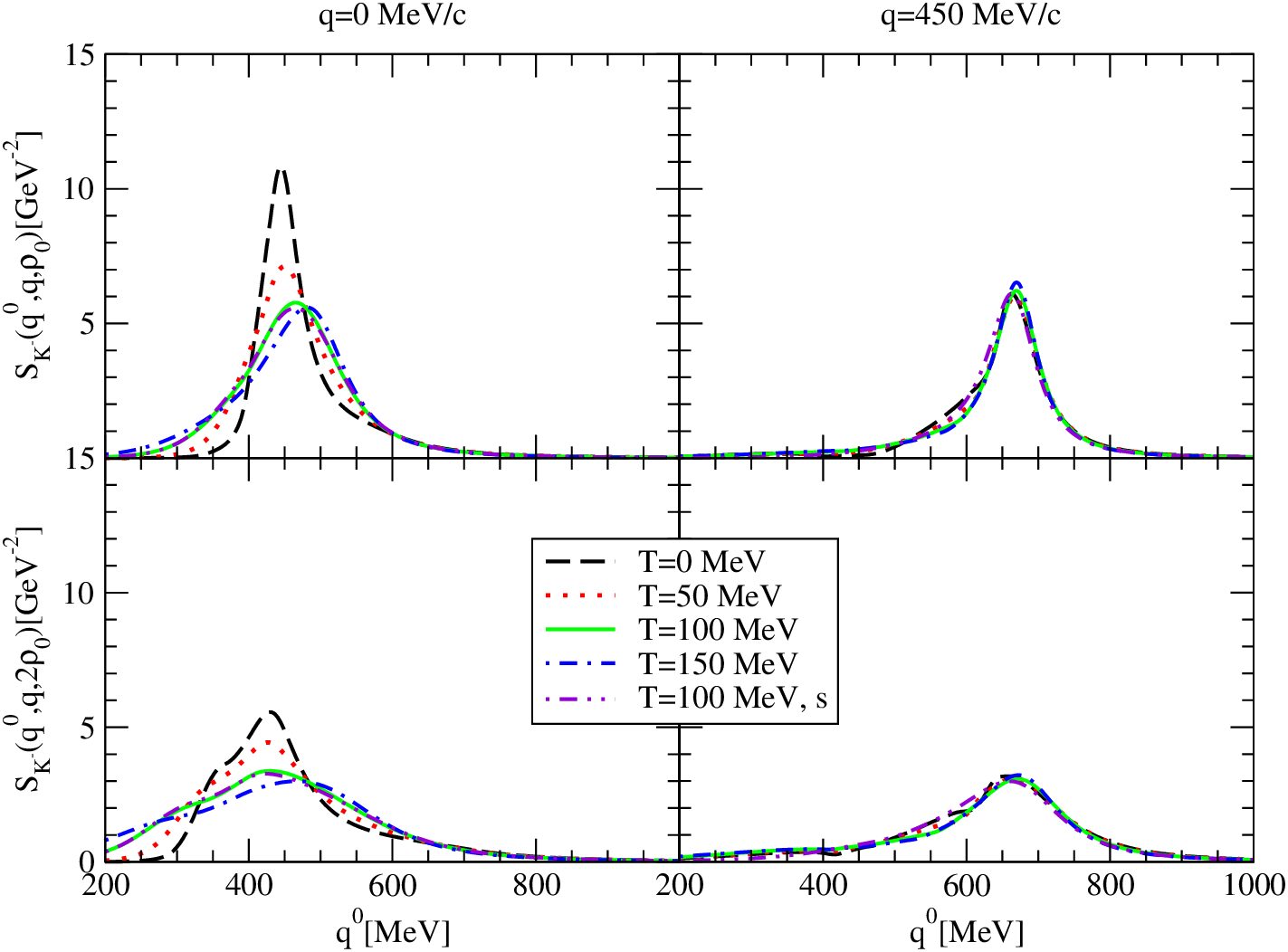}
\caption{Left: $\bar K$ self-energy and spectral function for $\rho=\rho_0$, $T=0,100$ MeV and two momenta showing the partial wave decomposition. Right: Evolution of the $\bar K$ spectral function with density and temperature for two momenta.}
 \label{fig1}
\end{center}
\end{figure}
%%%%%%%%%%%%%%%%%%%%%%%%%%%%%%%%%%%%%%%%%%%%%%%%%%%%

On the l.h.s. of Fig.~\ref{fig1} we show the real and imaginary parts of the $\bar K$ self-energy together with
the $\bar K$ spectral function as a
function of the $\bar K$ energy at $\rho_0=0.17 \ {\rm fm^{-3}}$ for two different momenta, $q=0$~MeV/c (left
column) and $q=450$~MeV/c (right column). The different curves correspond to
$T=0$ and $T=100$ MeV including the $s$-wave and the $(s+p)$-wave
contributions. The self-energy is dominated by the $s$-wave dynamics and the $p$-wave contributions from  $\Lambda N^{-1}$, $\Sigma N^{-1}$ and $\Sigma^*
N^{-1}$  excitations become evident at a finite momentum of $q=450$~MeV/c. The effect of these subthreshold excitations is repulsive at the $\bar K N$ threshold. This
repulsion together with the strength below threshold can be easily seen in the spectral function at finite momentum
(third row). At finite momentum the quasi-particle peak moves to higher energies while the spectral
function falls off slowly on the left-hand side. Temperature results in a softening of the real and imaginary part of the
self-energy as the Fermi surface is smeared out. The peak of the spectral
function moves closer to the free position while the spectral function extends over a wider range
of energies.

The evolution with density and temperature of the $\bar{K}$ spectral function is depicted 
on the r.h.s. of Fig.~\ref{fig1}. The spectral function shows a strong mixing between the quasi-particle peak and the $\Lambda(1405)N^{-1}$ and  $YN^{-1}$
excitations. As we have seen before, the effect of the $p$-wave $YN^{-1}$ subthreshold excitations is
repulsive for the $\bar K$ potential, compensating in part the attraction
from the $s$-wave ${\bar K} N$ interaction.  Temperature softens the
$p$-wave contributions to the spectral function at the quasi-particle energy. Moreover, together with the $s$-wave mechanisms, the $p$-wave self-energy
provides a low-energy tail which spreads the spectral function considerably.
 Increasing the density dilutes the spectral function even further.

\subsection{Energy weighted sum rules}

The hadron propagator or single particle Green's function has well defined analytical properties that impose some
constraints on the many-body formalism as well as the interaction model. An excellent tool to test the quality of our model for hadrons in medium is provided by the energy-weighted sum rules (EWSRs) of the single-particle spectral functions. The EWSRs are
obtained from matching the Dyson form of the meson propagator with its spectral Lehmann representation at low and high 
energies \cite{ewsr}. The first EWSRs in the high-energy limit expansion, $m_0^{(\mp)}$, together with the zero energy one, $m_{-1}$, are given by
\begin{eqnarray}
m_{-1}&:&
\int_0^{\infty} \textrm{d}\omega \,
\frac{1}{\omega} \, [ S_{\bar K}(\omega,\vec{q}\,;\rho,T) + S_{K}(\omega,\vec{q}\,;\rho,T)]
=
\frac{1}{\omega_{\bar K}^2(\vec{q}\,) + \Pi_{\bar K}(0,\vec{q}\,;\rho,T)} \ \ \ \
\\
m_{0}^{(\mp)}&:&  
\int_0^{\infty} \textrm{d}\omega \,
[ S_{\bar K}(\omega,\vec{q}\;\rho,T) - S_{K}(\omega,\vec{q}\,;\rho,T) ] = 0
\nonumber \\
& & 
\int_0^{\infty} \textrm{d}\omega \,
\omega \, [ S_{\bar K}(\omega,\vec{q}\,;\rho,T) + S_{K}(\omega,\vec{q}\,;\rho, T) ] = 1 \ .
%\nonumber \\
%\\
%m_{1}^{(\mp)}&:&  
%\int_0^{\infty} \textrm{d}\omega \,
%\omega^2 \, [ S_{\bar K}(\omega,\vec{q}\,;\rho,T) - S_{K}(\omega,\vec{q}\,;\rho,T) ] = 0
%\nonumber \\
%& & \ \ \ 
%\int_0^{\infty} \textrm{d}\omega \,
%\omega^3 \, [ S_{\bar K}(\omega,\vec{q}\,;\rho,T) + S_{K}(\omega,\vec{q}\,;\rho,T) ] 
%=
%\omega_{K}^2(\vec{q}\,) + \Pi_{\bar K}^{\infty}(\vec{q}\,;\rho,T) \ .
%\nonumber \\
\end{eqnarray}

%%%%%%%%%%%%%%%%%%%%%%%%%%%%%%%
\begin{figure}[t]
\begin{center}
\includegraphics[height=7 cm, width=6 cm]{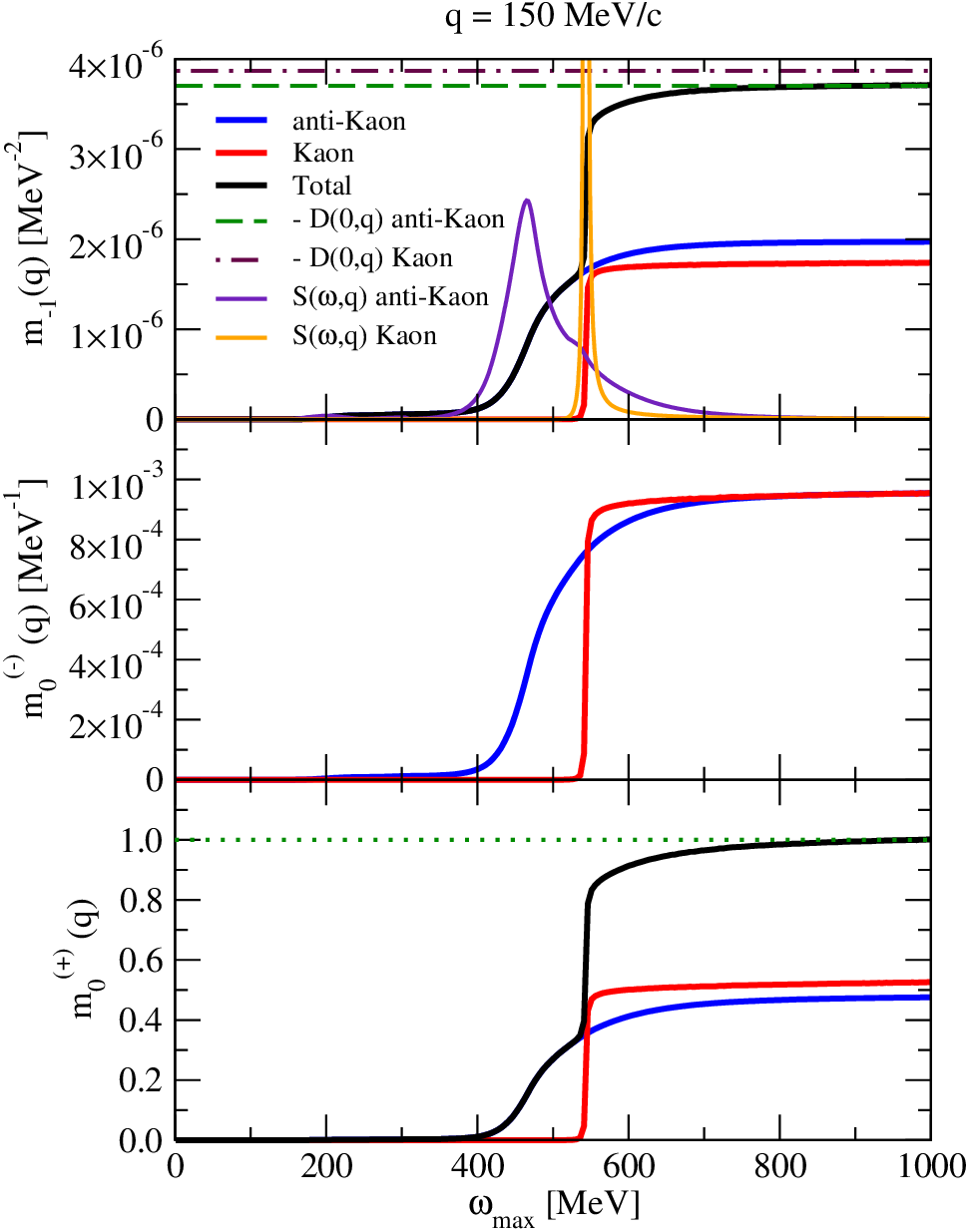}
%\includegraphics[height=6 cm, width=6 cm]{Sum-rule-KKbar-minusone-r0-T1-variousq-v2.eps}
%\hfill
%\includegraphics[height=5.8 cm, width=6 cm]{Sum-rule-KKbar-minusone-r0-T100-variousq-v2.eps}
\caption{ $m_{-1}$, $m_{0}^{(-)}$ and $m_{0}^{(+)}$ sum rules for the $K$ and $\bar K$ spectral functions at
$q=150$~MeV/c, $\rho=\rho_0$ and $T=0$ MeV. The $\bar K$ and $K$ spectral
functions are also displayed for reference in arbitrary units.}
 \label{fig2}
\end{center}
\end{figure}

%%%%%%%%%%%%%%%%%%%%%%%%%%%%%%%%%%%%%5

The sum rules for the antikaon propagator
are shown in Fig.~\ref{fig2} as a function of the upper integral limit in the case
of $\rho=\rho_0$, $T=0 \ {\rm MeV}$ and $q=150~$MeV/c. The contributions from $\bar K$ and $K$ to the l.h.s. of the sum rule
are depicted separately. The $\bar K$ and $K$ spectral functions are also shown for
reference in arbitrary units. Note that saturation is progressively shifted to higher energies as
we examine sum rules involving higher order weights in energy.

The l.h.s. of the $m_{-1}$ sum rule (upper panel) converges properly and saturates a few
hundred MeV beyond the quasiparticle peak, following the behaviour of the $\bar K$ and $K$ spectral functions. We have also plotted in
Fig.~\ref{fig2} the r.h.s. of the $m_{-1}$ sum
rule both for the antikaon and kaon, namely their off-shell propagators
evaluated at zero energy (modulo a minus sign).
The difference between both values reflects the violation of
crossing symmetry present in the chiral model employed
for the kaon and antikaon self-energies as we neglect the explicit
exchange of a meson-baryon pair in a $t$-channel configuration.
However, we  may still expect
the saturated value of the l.h.s. of the $m_{-1}$ sum-rule to provide a
constraint for the value of the zero-mode propagator appearing on the r.h.s, because
the most of the strength sets in at energies of the order of the meson
mass, where the  neglected  terms of the $K (\bar K)N$ amplitudes are
irrelevant.

The $m_0^{(-)}$ sum rule shows that the areas subtended by the $K$ and $\bar
K$ spectral functions should coincide. This is indeed the case for the
calculation considered here, as can be seen in the middle panel of
Fig.~\ref{fig2}. The fullfilment of this sum rule is, however, far from trivial.
We recall that whereas one expects the $\bar K$ and $K$ spectral
functions to be related by the retardation property, $S_{\bar
K}(-\omega)=-S_K(\omega)$, the actual calculation of the meson self-energies is
done exclusively for positive meson energies.

The $m_0^{(+)}$ sum rule saturates to one independently of the meson momentum,
nuclear density or temperature, thus posing a strong constraint on the accuracy
of the calculations.
The lower panel in Fig.~\ref{fig2} shows that the $K$ and
$\bar K$ spectral functions fulfill this sum rule to a high
precision.

We have also tested those sum rules for higher momenta and temperature.
As the meson momentum is increased, the saturation
of the integral part of the sum rules is progressively shifted to higher
energies, following the strength of the spectral distribution. At finite temperature
the $\bar K$ spectral function spreads considerably \cite{Tolos:2008di}, and
in particular acquires a sizable low energy tail from smearing of the Fermi
surface, which contributes substantially to the l.h.s. of the sum rule below
the quasi-particle peak. The $K$ contribution also
softens at finite temperature and increasing momenta, as the $K$ in-medium decay width is
driven by the $KN$ thermal phase space.

\section{Charm mesons}

The properties of charm mesons is a topic of recent analysis and lot of effort is being invested in constructing effective models for the meson-baryon interaction in the charm sector. In this section we present two different approaches to this effective meson-baryon interaction. We then discuss possible experimental scenarios where the charm meson properties can be tested, such as scalar resonances in nuclear matter.

\subsection{SU(4) $t$-vector meson exchange models}
\label{su4}

The $D$ and $\bar D$  meson spectral functions are obtained from the multichannel Bethe-Salpeter equation taking, as bare interaction, a type of broken $SU(4)$ $s$-wave WT  interaction supplemented by an attractive isoscalar-scalar term and using a cutoff regularization scheme. This cutoff is fixed by the position and the width of the $I=0$ $\Lambda_c(2593)$ resonance. As a result, a new resonance in $I=1$ channel $\Sigma_c(2880)$ is generated \cite{LUT06}. The in-medium solution at finite temperature 
incorporates, as well, Pauli blocking effects, baryon mean-field bindings and $\pi$ and $D$ meson self-energies \cite{TOL07}.

\begin{figure}[t]
\begin{center}
\includegraphics[height=7 cm, width=6 cm]{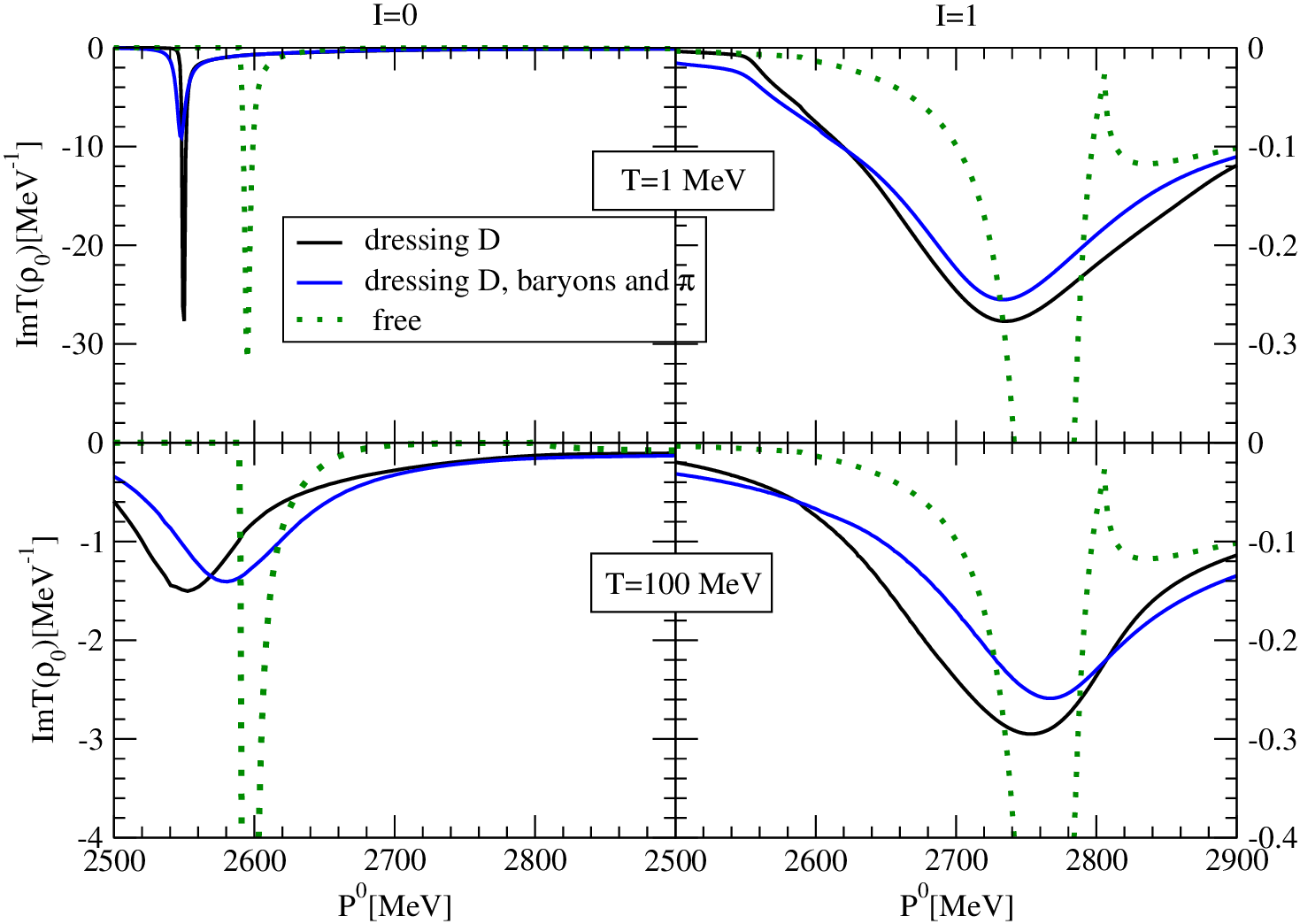}
\hfill
\includegraphics[height=7 cm, width=6 cm]{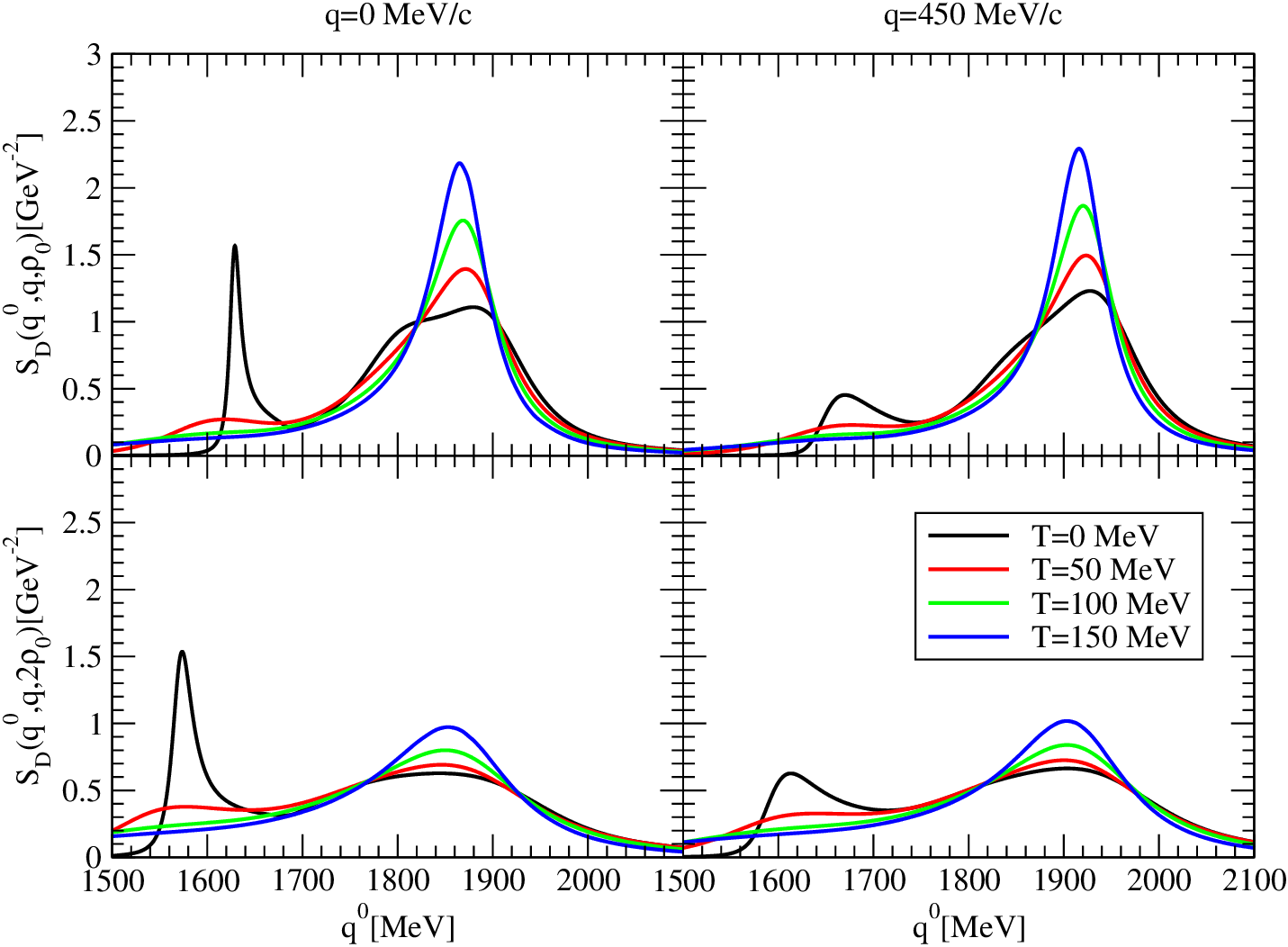}
\caption{Left:$\tilde{\Lambda}_c$ and $\tilde{\Sigma}_c$ resonances. Right: The $D$ meson spectral function} \label{fig3}
\end{center}
\end{figure}

The $I=0$ $\tilde{\Lambda}_c$ and $I=1$ $\tilde{\Sigma}_c$ resonances in hot dense matter are shown in the l.h.s. of Fig.~\ref{fig3} for two different self-consistent calculations:  i) including only the
self-consistent dressing of the $D$ meson,  ii) 
including mean-field binding on baryons and the pion self-energy. Medium effects at $T=0$ lower the
position of the $\tilde{\Lambda}_c$ and $\tilde{\Sigma}_c$ with respect
to their free  values. Their width values, which increase due to $\tilde
Y_c (=\tilde \Lambda_c, \tilde \Sigma_c) N \rightarrow \pi N \Lambda_c, \pi N \Sigma_c$  processes, differ according to the phase space available. The pion dressing induces a small effect in the resonances because of charm-exchange channels being suppressed. Finite temperature results in the reduction of the Pauli blocking due to the smearing of the Fermi surface. Both resonances move up in energy closer to their free position while they are smoothed out, as in \cite{Tolos:2006ny}.

%%%%%%%%%%%%%%%%%%%%%%%%%%%%%%%%%%%%%%%%%%%%%%%%%%%%%%%%%%%%%%%%%%%%%%%%%%%%5
%%%%%%%%%%%%%%%%%%%%%%%%%%%%%%%%%%%%%%%%%%%%%%%%%%%%%%%%%%%%%%%%%%%%%%%%%%%%%

In the r.h.s of Fig.~\ref{fig3} we display the evolution with density and temperature of the $D$ meson spectral function for (ii). At $T=0$ the spectral function presents two peaks: $\tilde \Lambda_c N^{-1}$ excitation at a lower energy whereas the second one
at higher energy is the quasi(D)-particle  peak  mixed with  the $\tilde \Sigma_c N^{-1}$ state.  Those structures dilute with increasing temperature while the quasiparticle peak gets closer to its free value becoming narrower, as the self-energy
receives contributions from higher momentum $DN$ pairs where the interaction is weaker.
Finite density results in a broadening of the spectral function because of the increased decay and collisional phase space. Similar effects were observed previously for the $\bar K$ in hot dense nuclear matter \cite{Tolos:2008di}.

\subsection{SU(8) scheme with heavy-quark symmetry}

%%%%%%%%%%%%%%%%%%%%%%%%%%%%%%%
\begin{figure}[t]
\begin{center}
\includegraphics[height=7 cm, width=6 cm]{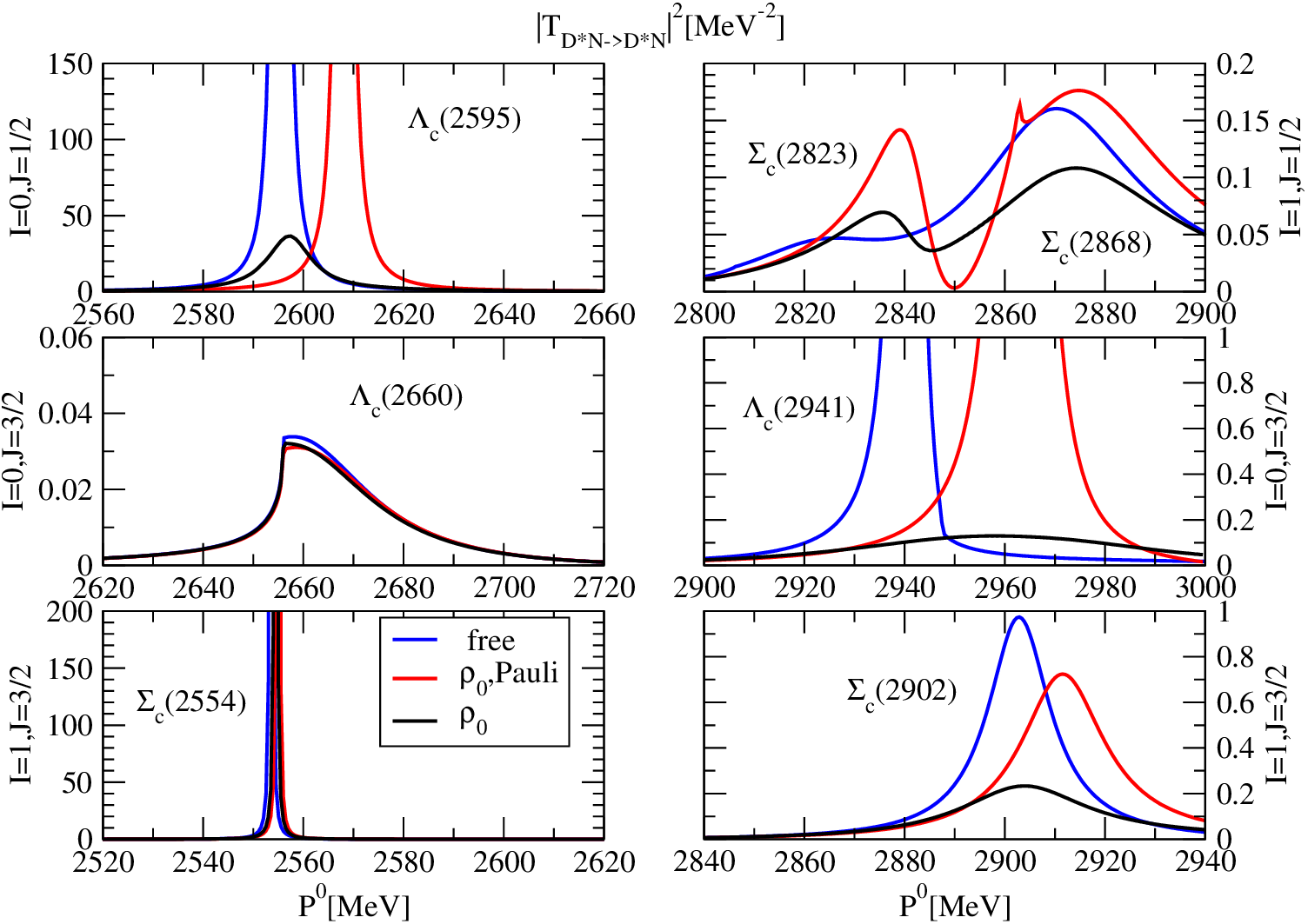}
\hfill
\includegraphics[height=6 cm, width=6 cm]{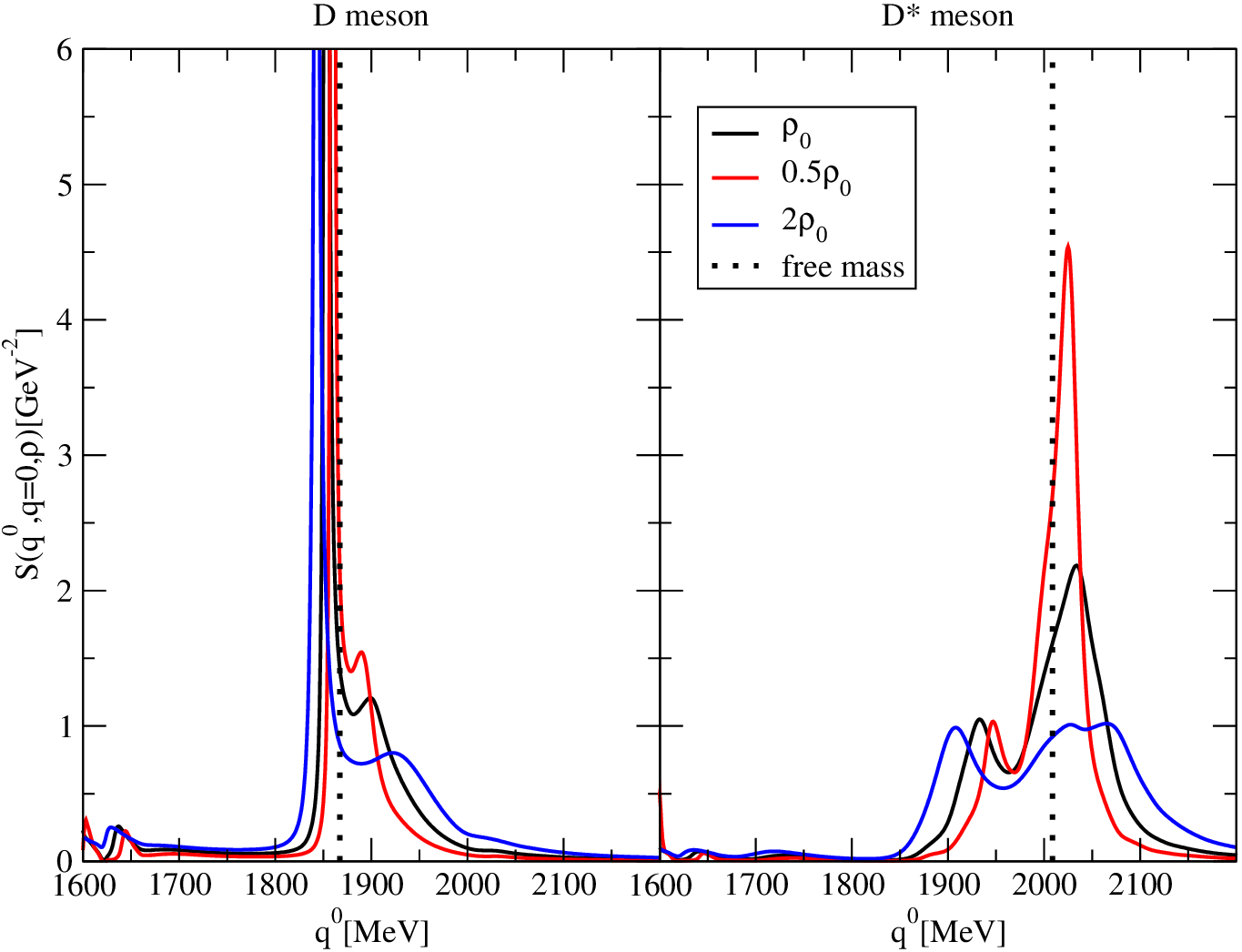}
\caption{Left: Dynamically-generated charmed baryonic resonances in nuclear matter. Right: $D$ and $D^*$ spectral functions in nuclear matter at $q=0$ MeV/c.} \label{fig4}
\end{center}
\end{figure}

%%%%%%%%%%%%%%%%%%%%%%%%%%%%%%%%%%%%%5

Heavy-quark symmetry (HQS) is a proper QCD spin-flavor symmetry
that appears when the quark masses, such as the charm mass, become
larger than the typical confinement scale. As a consequence of this
symmetry, the spin interactions vanish for infinitely massive
quarks. Thus, heavy hadrons come in doublets (if the spin of the light
degrees of freedom is not zero), which are degenerated in the infinite
quark-mass limit. And this is the case for the $D$ meson and its
vector partner, the $D^*$ meson.

Therefore, we calculate the self-energy and, hence, the spectral function of the $D$ and $D^*$ mesons in nuclear matter simultaneously from a self-consistent calculation in coupled channels. To incorporate HQS to the meson-baryon interaction
we extend the WT meson-baryon lagrangian to the $SU(8)$ spin-flavor
symmetry group as we include pseudoscalars and vector mesons together with $J=1/2^+$ and $J=3/2^+$ baryons \cite{magas}, following the steps for $SU(6)$ of Ref.\cite{GarciaRecio:2005hy}. However, the $SU(8)$ spin-flavor is strongly broken in nature. On one hand, we take into account mass
breaking effects by adopting the physical hadron masses in the tree
level interactions and in the evaluation of the
kinematical thresholds of different channels, as done in $SU(4)$ models. On the other hand, we consider the difference between the weak non-charmed and charmed
pseudoscalar and vector meson decay constants. We also improve on the regularization scheme in nuclear matter going beyond the usual cutoff scheme \cite{tolos09}.

The $SU(8)$ model generates a wider spectrum of resonances with charm $C=1$
and strangeness $S=0$ content compared to the previous $SU(4)$ models, as seen in the l.h.s of Fig.~\ref{fig4}. While the parameters of both $SU(4)$ and $SU(8)$ models are fixed by the ($I=0$,$J=1/2$)
$\Lambda_c(2595)$ resonance, the incorporation of vectors mesons in
the $SU(8)$ scheme generates naturally $J=3/2$ resonances, such as
$\Lambda_c(2660)$, $\Lambda_c(2941)$, $\Sigma_c(2554)$ and
$\Sigma_c(2902)$, which might be identified experimentally
\cite{Amsler}. New resonances are also produced for $J=1/2$, as
$\Sigma_c(2823)$ and $\Sigma_c(2868)$, while others are not observed in $SU(4)$ models due to the different symmetry breaking pattern used in both
models. The modifications of the mass and width of these resonances in
the nuclear medium will strongly depend on the coupling to channels
with $D$, $D^*$ and nucleon content. Moreover, the resonances close to
the $DN$ or $D^*N$ thresholds change their properties more evidently
as compared to those far offshell. The improvement in the
regularization/renormalization procedure of the intermediate
propagators in the nuclear medium beyond the usual cutoff method has
also an important effect on the in-medium changes of the
dynamically-generated resonances, in particular, for those lying far
offshell from their dominant channel, as the case of the
$\Lambda_c(2595)$.

%%%%%%%%%%%%%%%%%%%%%%%%%%%%%%%
\begin{figure}[t]
\begin{center}
\includegraphics[height=6 cm, width=6 cm]{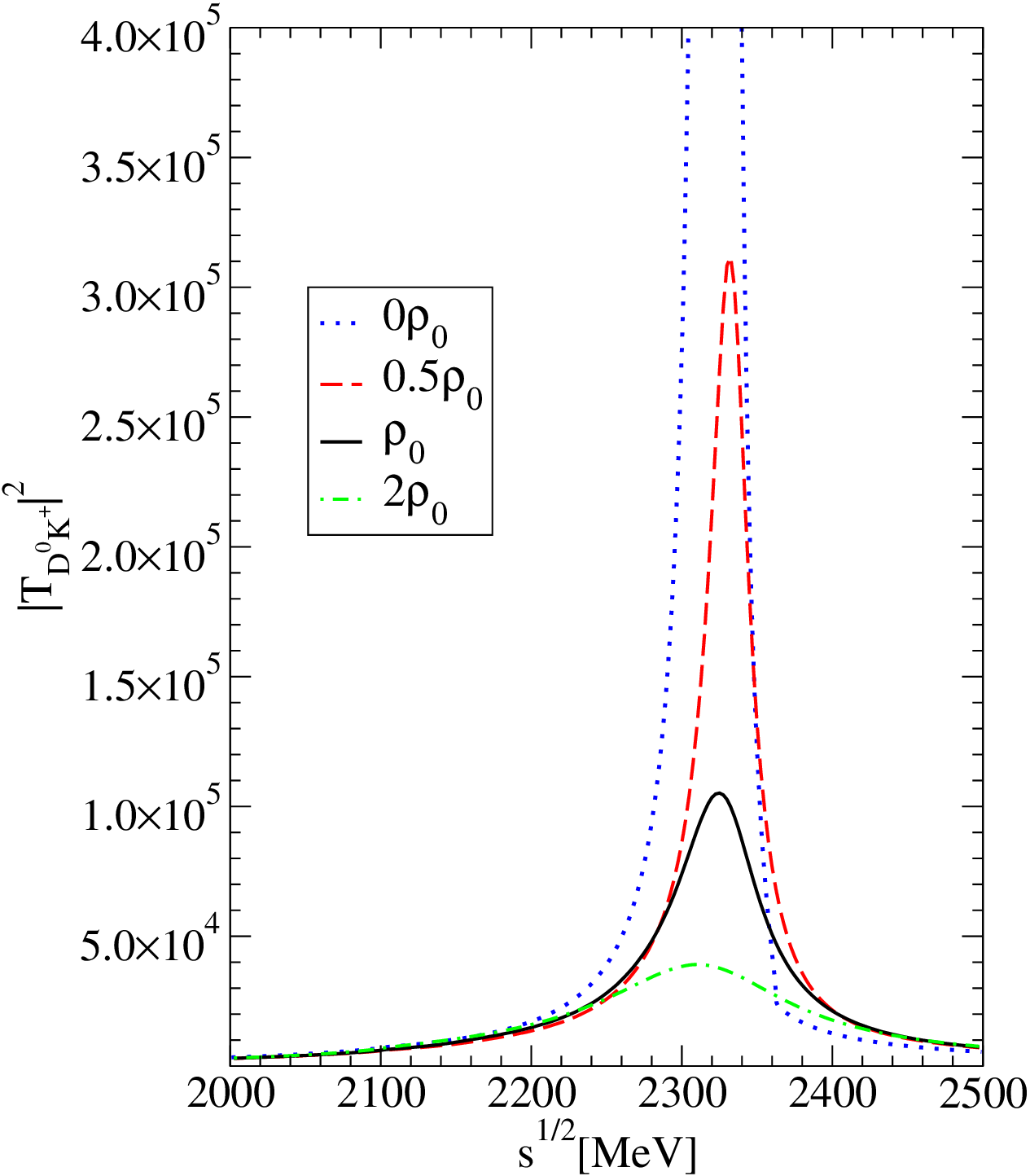}
\hfill
\includegraphics[height=6 cm, width=6 cm]{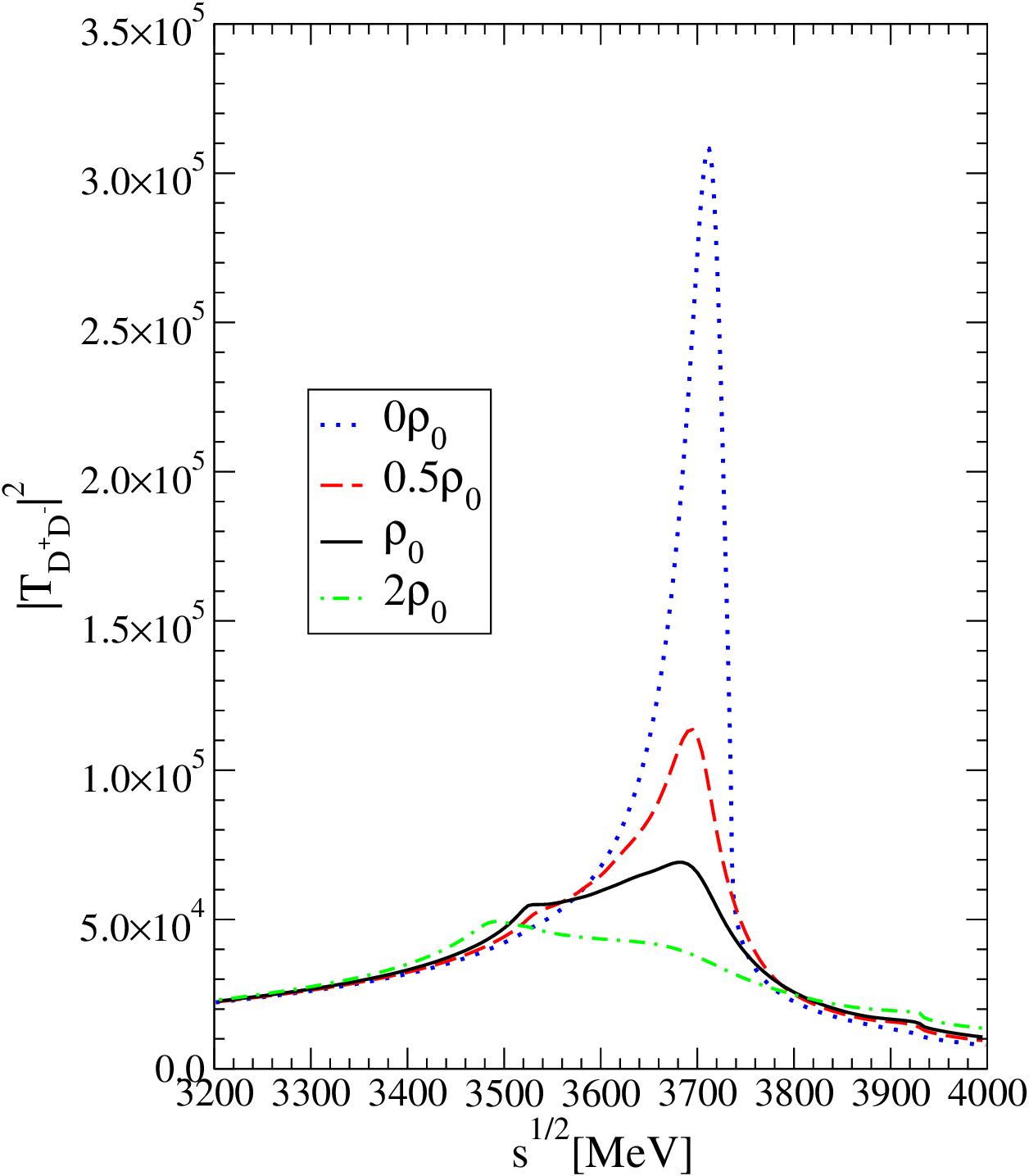}
\caption{$D_{s0}(2317)$ (left) and $X(3700)$ (right) resonances.}
\label{fig5}
\end{center}
\end{figure}

%%%%%%%%%%%%%%%%%%%%%%%%%%%%%%%%%%%%%5

In the r.h.s of Fig.~\ref{fig4} we also display the $D$ and $D^*$ spectral functions, which show then a rich spectrum of resonant-hole states. The $D$
meson quasiparticle peak mixes strongly with $\Sigma_c(2823)N^{-1}$
and $\Sigma_c(2868)N^{-1}$ states while the $\Lambda_c(2595)N^{-1}$ is
clearly visible in the low-energy tail. The $D^*$ spectral function
incorporates the $J=3/2$ resonances, and the quasiparticle peak fully mixes with $\Sigma_c(2902)N^{-1}$ and $\Lambda_c(2941)N^{-1}$. As
density increases, these $Y_cN^{-1}$ modes tend to smear out and the
spectral functions broaden as the collisional and absorption processes
increase.

\subsection{Charm and hidden charm resonances in nuclear matter}

The nature of a resonance, whether it has the usual $q \bar q$/$qqq$ structure or is better described as being dynamically generated, is an active matter of research, in particular, for scalar resonances. The excitation mechanisms in the nucleus together with the properties of those particles  can be extracted studying their renormalized properties in nuclear matter.

We study the charmed resonance $D_{s0}(2317)$ \cite{Kolomeitsev:2003ac,Gamermann:2006nm} together with a hidden charm scalar meson, $X(3700)$, predicted in \cite{Gamermann:2006nm}, which might have been observed by the Belle collaboration \cite{Abe:2007sy} via the reanalysis of \cite{Gamermann:2007mu}. Those resonances are generated dynamically solving the coupled-channel Bethe-Salpeter equation for two pseudoscalars \cite{Molina:2008nh}. The kernel is derived from a $SU(4)$ extension of the $SU(3)$ chiral Lagrangian used to generate scalar resonances in the light sector. The $SU(4)$ symmetry is, however, strongly 
 broken, mostly due to the explicit consideration of the masses of the vector 
 mesons exchanged between pseudoscalars \cite{Gamermann:2006nm}. 

The analysis of the transition amplitude close to each resonance for the different coupled channels gives us information about the coupling of the resonance to a particular channel. The $D_{s0}(2317)$ mainly couples to the $DK$ system, while the hidden charm state $X(3700)$ couples most strongly to $D\bar{D}$. Therefore, any change in the $D$ meson properties in nuclear matter will have an important effect on these  resonances. Those modifications are given by the $D$ meson self-energy, as discussed in Sec.~\ref{su4}, but supplemented by the $p$-wave self-energy through the corresponding $Y_cN^{-1}$ excitations \cite{Molina:2008nh}.

 In Fig.~\ref{fig5}, the resonances $D_{s0}(2317)$ and $X(3700)$ are shown by displaying the squared transition amplitude for the corresponding dominant channel at different densities. The $D_{s0}(2317)$ and 
$X(3700)$ resonances, which have a zero and small width, respectively,
develop widths of the order of 100 and 200
MeV at normal nuclear matter density, correspondingly. The origin can be traced back to the opening of new many-body decay channels, as the $D$ meson gets absorbed in the nuclear medium via $DN$ and $DNN$ inelastic reactions. In our model, we do not extract any clear conclusion for the mass shift. We suggest to look at transparency ratios to investigate those in-medium widths. This magnitude, which gives the survival probability in production reactions in  nuclei, is very sensitive to the absorption rate of any resonance inside nuclei, i.e., to its in-medium width.
   
\section{Conclusions and Outlook}

In this paper we have studied the properties of strange and charm mesons in hot and dense matter within a self-consistent coupled-channel approach. The in-medium solution at finite temperature accounts for Pauli blocking effects, mean-field binding on all the baryons involved, and meson self-energies. We have analyzed the behaviour in this hot and dense environment of dynamically-generated baryonic resonances together with the evolution with density and temperature of the strange and open-charm meson spectral functions. The spectral function for $\bar K$ and $D$ mesons dilutes with increasing temperature and density while the quasiparticle peak moves closer to the free position.
The spectral function for strange mesons is also tested using energy-weighted sum rules and we found that the sum rules for the lower energy weights are fulfilled satisfactorily. We have finally discussed the implications of the properties of charm mesons on the  $D_{s0}(2317)$ and the predicted $X(3700)$. We suggest to look at transparency ratios to investigate the changes in width of those resonances in nuclear matter.

\section{Acknowledgments}
L.T. acknowledges support from the RFF-Open and Hidden Charm at PANDA project of the University of Groningen. This work is partly supported by the EU contract No. MRTN-CT-2006-035482 (FLAVIAnet), by the contracts FIS2006-03438, FIS2008-01661, FIS2008-01143 and FPA2009-00592 from MICINN (Spain), by the Spanish Consolider-Ingenio 2010 Programme CPAN (CSD2007-00042), by the Generalitat de Catalunya contract 2009SGR-1289, by the UCM-BSCH contract GR58/08 910309 contracts and by Junta de Andaluc\'{\i}a under contract FQM225. We acknowledge the support of the European Community-Research Infrastructure Integrating Activity ``Study of Strongly Interacting Matter'' (HadronPhysics2, Grant Agreement n. 227431) under the 7th Framework Programme of EU.

\end{document}